\documentclass[conference]{IEEEtran}
\usepackage{graphicx}
\usepackage{amsfonts,amssymb}
\usepackage{lipsum}
\usepackage{bm}
\usepackage{epstopdf}
\usepackage{subeqnarray}
\usepackage{algorithm}
\usepackage{float}
\usepackage{algorithmic}
\usepackage{subfigure}
\usepackage{tabularx} 
\usepackage{amsmath}
\usepackage{booktabs }
\usepackage{amsthm} 
\usepackage{cite}
\usepackage{multirow}
\graphicspath{{figures/}}
\begin{document}

\title{ Distributed Private Online Learning for Social Big Data Computing over Data  Center Networks }

\author{\IEEEauthorblockN{Chencheng Li$^1$, Pan Zhou$^2$\dag, \emph{Member, IEEE}, Yingxue Zhou$^3$, Kaigui Bian$^4$, \emph{Member, IEEE}, \\Tao Jiang$^5$, \emph{Senior Member, IEEE}, Susanto Rahardja$^6$, \emph{Fellow, IEEE}\\}
\IEEEauthorblockA{School of EIC, Huazhong University of Science \& Technology, Wuhan, China$^{{1,2,3,5}}$\\School of EECS, Peking University$^4$\\ Northwestern Polytechnical University and National University of Singapore$^6$ \\
Email: lichencheng@hust.edu.cn$^1$, panzhou@hust.edu.cn$^2$, \dag Corresponding author of this paper} }
\maketitle

\begin{abstract}
With the rapid growth of Internet technologies, cloud computing and social networks have become ubiquitous.   An increasing number of people  participate in social networks and massive online social data are obtained. In order to exploit knowledge from copious amounts of data obtained and predict social behavior of users, we urge to realize data mining in social networks. Almost all online websites use cloud services to effectively process the  large scale of social data, which are gathered from distributed data centers. These data are so large-scale, high-dimension and widely distributed that we propose a distributed sparse online algorithm to handle them. Additionally, privacy-protection is an important point in social networks. We should not compromise  the privacy of individuals in networks, while these social data are being learned for data mining. Thus we also  consider the privacy problem in this article. Our simulations shows that the appropriate sparsity of data would enhance the  performance of our algorithm and the privacy-preserving method does not significantly hurt the  performance of the proposed algorithm.

\end{abstract}


\IEEEpeerreviewmaketitle

\section{Introduction}
A social network is referred as a structure of  ``Internet users''  interconnected through a variety of relations \cite{ahuja2013}. For a single user, he/she has some different relationships in different social networks such as friends and followers. Also, one user  has diverse social activities, e.g., post messages, photos and other media on Facebook and upload, view, share and comment on videos on YouTube.  According to statistics, almost $500$ TB social data are generated per day. It takes high operational costs to store the data and it is a waste of resources without using them. Hence, we want to conduct the social big data analysis, in which the users  active in a social, collaborative context to make sense of data.  However, handling such a volume of social data brings us many challenges. We next describe the main challenges and the corresponding approaches to them.

The social data are generated all around the world and collected over distributed sources into different and interconnected data centers. Hence, it is hard to process the data in a centralized model. Concerned with this problem, cloud computing may be a good choice.  As is known, many social networking websites (e.g., Facebook, Twitter, LinkedIn and YouTube) obtain computing resources across a network. These corporations host their social networks  on a cloud platform. This cloud-based model owns some advantages, chief among which is the lowered costs in infrastructure.  They can rent cloud computing services from other third part due to their actual needs and scale up and down at any time without taking additional cost in infrastructure \cite{dusenbery2012}. Beyond that, they are able to choose different cloud computing services according to the distribution of social data. Naturally, for social data analysis in cloud, a distributed online learning algorithm  is needed to handle the massive social data in distributed scenarios \cite{jiang2014}. Based on cloud computing, we equip each data center with the independent online learning ability and they can exchange information with other data centers across the network. Each data center is urged to build a reliable model to recommend its local users without directly sharing social data with each other.
In theory, this approach is a distributed optimization technology and many researches \cite{duchi2012dual,nedic2009distributed,ram2010distributed} have been devoted to it.  To estimate the utility of the proposed model, we use the notion ``regret'' \cite{shalev2011} in online learning (see Definition 3).

In Big Data era, social big data are both large scale and  high dimension. A single person has a variety of social activities in a social network, so the corresponding vector of his/her social information is ``long''. However, when a data miner studies the  consumer behavior about one interest, some of the  information  in the vector  may not be relevant. For example, a person's height and age cannot contribute to predicting his taste. Thus, high dimension could enhance the  computational complexity of  algorithms and weaken the utility of online learning models. To deal with this problem, we introduce a sparse solution in social big data.  In this paper, we introduce two classical groups of effective methods for sparse online learning \cite{wang2013large, wang2015, shalev2011b}. The first group (e.g., \cite{langford2009}) induces sparsity in the weights of online learning algorithms via truncated gradient. The second group studies on sparse online learning follows the dual averaging algorithm \cite{xiao2010}. In this paper, we will exploit \emph{online mirror descent} \cite{duchi2010} and  \emph{Lasso-$L_1$ norm}  \cite{tibshirani1996} to make the parameter updated in algorithm sparse.

Furthermore, exchanging information contained in social data among data centers may lead to  privacy breaches  as it  flows across the social network.  Once social data are mined without any security precautions,  it is of high probability to divulge privacy.
Admittedly, preserving privacy consequentially lead to the lowered performance of knowledge discovery in cloud-based social data.
Therefore, we intend to design an algorithm, which protects the privacy while makes full use of the social data. Finally, we choose the ``differential privacy'' \cite{dwork2006} technology to guarantee the safety of data centers in cloud. At a high level, a differentially private online learning model guarantees that its output of data mining does not change ``too much"  because of perturbations (i.e., add some random noise to the data transmitted) in any individual social data point. That means whether or not a data point  being in the database, the mining outputs are difficult to distinguish and then the miner cannot obtain the sensitive information based on search results. 

In conclusion, we make three contributions: 1) we propose a distributed online learning algorithm to handle decentralized social data in real time and demonstrate its feasibility; 2)  sparsity is induced to compute the high-dimension social data for enhancing the accuracy of predictions; 3) differential privacy is used to protect the privacy of data without seriously weaken the performance of the online learning algorithm.

This paper is organized as follows. Section II introduces the system model and propose the algorithm. The privacy analysis is done in Section III.  We analyze the utility of the algorithm in Section IV. Numerical results and performance improvements are shown in Section V. Section VI concludes the paper.

 \section{system model}
In this section, the system model and our private online learning algorithm are presented. 

Consider a social network, in which all online users are served on cloud platforms, e.g., Fig.1.  These users operate on their own personal page and the generated social data are collected and transmitted to the nearest data center on cloud, just as shown in Fig.1, all data are collected by the data centers marked with $A \to G$. Because of the huge network, many data centers are widely distributed. Each data center has its corresponding cloud computing node, where the nearby social data are processed in  real time . As a holonomic  system, the social network should have a good knowledge of  all data it owns, thus data centers should exchange information with each other.  Since there are too many  data centers and  most of them are located over the world, a data center never can communicate with all other centers. To achieve better economic benefits, each data center just can exchange information with neighboring ones (e.g., $D$  is just connected to its adjacent centers $C$ and $G$).  Furthermore, random noise should be added to each communication for protecting the  privacy (yellow arrows in Fig.1). Since such social big data need to be efficiently and privately processed with the limited communications,  we focus on distributed optimization  and differential privacy technologies.

We next introduce how the communications among data centers on cloud are conducted. Recall that we intend to realize knowledge discovery in social data in real time. A  new parameter, e.g., $w$, should be created to denote the online learning  parameter (containing the knowledge mined from data). At each iteration, each cloud node updates $w$ based on its local data center and then exchanges $w$ with neighbors. This communication mechanism forms a network topology. The network topology can be fixed or time-variant, which is proved  to have no great influence on the utility of our algorithm in Section IV. 

 \subsection{Communication Graph}

  For our online learning social network, we denote the communication matrix by $A$ and let $a_{ij}$ be the $(i,j)$-th element of $A$. In the system, $a_{ij}$ is the weight of  the learning parameter which the $i$-th cloud node transmits to the $j$-th one. ${a_{ij}}(t)>0$ means there exists a communication between the $i$-th and $j$-th nodes at round $t$, while ${a_{ij}}(t)=0$  means non-communication between them. . For a clear description, we denote the communication graph for a node $i$ at round $t$ by 
\begin{eqnarray}{\mathcal{G}_i} = \{ (i,j):{a_{ij}} > 0\},\end{eqnarray}
 where $\notag{a_{ij}}\in {A}.$
 
 To achieve the global convergence, we make some assumptions about $A$.
 
  \begin{figure}[tbp]
  \label{fig:subfig:a}
\includegraphics[width=3.5in]{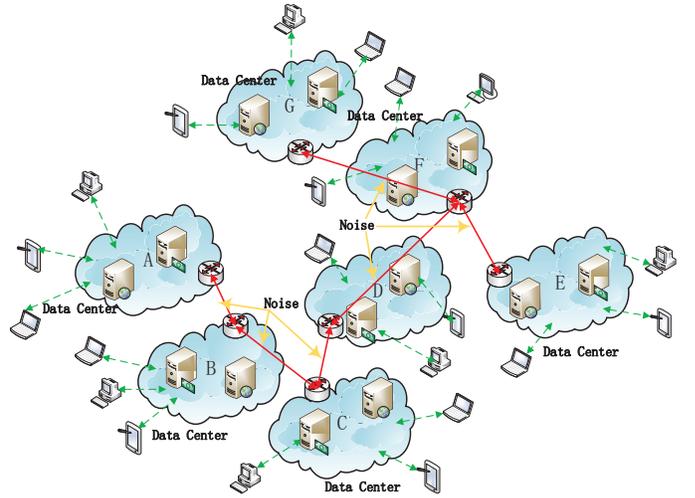}
\caption{Private Social Big
Data Computing over Data Center Networks
} 
\vspace{-.9em}
\end{figure}

\textbf{Assumption 1.} For an arbitrary node $i$, there exists a minimal scalar $\eta$, $0 < \eta  < 1$ such that
\begin{itemize}
\item[(1)]${a_{ij}} > 0 $ for $(i,j) \in \mathcal{G}_i$,
\item[(2)] $\sum\nolimits_{j = 1}^m {{a_{ij}}}  = 1$ and $\sum\nolimits_{i = 1}^m {{a_{ij}}}  = 1$,
\item[(3)] ${a_{ij}} > 0$ implies that  ${a_{ij}}\ge \eta $.
\end{itemize}

Here, Assumptions (1) and (2) state that each node computes a weighted average of neighboring learning parameters. Assumption (3) ensures that the influences among the nodes are significant. 

The above assumption is a necessary condition which presents in all researches (e.g.,\cite{duchi2012dual,nedic2009distributed,ram2010distributed}) about distributed optimization. Fortunately, this technology can be used to solve our distributed online learning in social networks.

 \subsection{Sparse Online Learning}
 
   As described, a social data is high dimensional. Hence, its corresponding learning parameter $w$ is a  long vector. In order to find the factors most related to one predicting behavior, we need to aggressively make the irrelevant  dimensions zero. Lasso \cite{tibshirani1996} is a famous method to produce some coefficients that are exactly $0$. Lasso cannot be directly used in the algorithm, we combine it with online mirror descent (see Algorithm 1) which is a special online learning algorithm.

For convenient analysis, we next  make some assumptions about the mathematical model of online learning system in social network. We assume to have $m$ data centers over the social network. Each data center collects massive social data every minute and processes them on cloud computing. For the data generated from social networks, we use $x$ to denote the social data of individual person. $\widehat y$ (e.g., $\widehat y = {w^T}x$)
 denotes the prediction for a  user, which helps the online website offer the user satisfying service. Then, the user will give a feedback denoted by $y$ telling the website whether the previous  prediction makes sense for him. Finally, due to the loss function (e.g., $f\left( {w,x,y} \right) = {\left[ {1 - {w^T}x \cdot y} \right]_ + }$), we compare the $\widehat y$ and  $y$ to find how many ``mistakes'' the online learning algorithm makes. Summing these ``mistakes'' over time and social networks, we obtain the regret of the whole system, based on which we can know the performance of our algorithm. 
 
\textbf{Assumption 2.} Let $W$ denote the set of $w$, we assume $W$ and the loss function $f$ satisfy:
\begin{itemize}
\item[(1)] The set $W$ is closed and convex subset of $\mathbb{R}^n$. Let $R \buildrel \Delta \over = \mathop {\sup }\limits_{x,y \in W} \left\| {x - y} \right\|$ denotes the diameter of $W$.
\item[(2)] The loss function $f$ is \emph{strongly convex} with modulus $\gamma  \ge 0$. For all $x,y \in W$, we have
\vspace{-.5em}\begin{eqnarray}
\left\langle {\nabla f_t^i,y - x} \right\rangle  \le f_t^i(y) - f_t^i(x) - \frac{\gamma }{2}{\left\| {y - x} \right\|^2}.
\end{eqnarray}
\item[(3)] The subgradients of $f$ are uniformly bounded, i.e., there exists $L > 0$ , for all $x \in W$, we have
\vspace{-.5em}\begin{eqnarray}\left\| {\nabla f_t^i(x)} \right\| \le L.\end{eqnarray}
\end{itemize}

Assumption (1) guarantees that there exists an optimal solution in our algorithm. Assumptions (2) and (3) help us analyze the convergence of our algorithm.

 \subsection{Differential Privacy} Dwork \cite{dwork2006} first proposed the definition of differential privacy which makes a data miner be able to release some statistic of its database without revealing sensitive information about a particular value itself. In this paper, we realize  output perturbation by adding a random noise denoted by $\delta$. This noise interferes some malicious data miners to steal sensitive information (e.g., birthday and contact info). Based on the parameters defined above, we give the following definition.
 
\textbf{Definition 1.} Assume that $\mathcal{A}$ denotes our differentially private online learning algorithm. Let $\mathcal{X}=\left\langle {x_1,x_2,...,x_T} \right\rangle $ be a sequence of social data taken from an arbitrary node's  local data center. Let $\mathcal{\theta } = \left\langle {\theta _1,\theta _2,...,\theta_T} \right\rangle $ be a sequence of $T$ results of  the node and $\mathcal{\theta} =\mathcal{A}(\mathcal{X})$. Then, our algorithm $\mathcal{A}$ is $\epsilon$-differentially private if given any two adjacent question sequences $\mathcal{X}$ and $\mathcal{{X'}}$ that differ in one social data entry, the following holds: 
 \begin{eqnarray}\Pr \left[ {\mathcal{A\left( X \right)}} \right] \le {e^\epsilon}\Pr \left[ {\mathcal{A\left( {X'} \right)}} \right].\end{eqnarray} 

This  inequality guarantees that whether or not an individual participates in the database, it will not make any significant difference on the output of our algorithm, so the adversary is not able to gain useful information about the individual person.

\subsection{Private Distributed Online Learning Algorithm}
We present a private distributed online learning algorithm for cloud-based social networks. Specifically, each cloud computing node propagates the parameter with noise added to neighboring nodes. After receiving the parameters from others, each node compute a weight average of the received and its old parameters. Then, each node updates the parameter due to general online mirror descent and induce sparsity using Lasso. The algorithm is summarized in Algorithm 1. Note that $w_t^i$ denotes the parameter of the $i$-th cloud node at time $t$. ${\varphi _{t{\rm{ = 1,}}...{\rm{,T}}}}$ are a series of  $\beta$-strongly convex functions.

\begin{algorithm}
\caption{Private Distributed Online Learning}
\begin{algorithmic}[1]
\STATE \textbf{Input}: Cost functions $f_t^i(w ): = \ell (w,x_t^i,y_t^i)$, $i \in [1,m]$ and $t \in [1,T]$; double stochastic matrix $A = (a_{ij}) \in {R^{m \times m}}$; 
\STATE \textbf{Initiaization:} $\theta _1^i=0$, $i \in [1,m]$ 
\FOR {$t = 1,...,T$ }
 \FOR{each node $i = 1,...,m$}
\STATE receive $x_t^i \in {\mathbb{R}^n}$
\STATE  $p_t^i = \nabla \varphi _t^ * \left( {{\theta _t^i}} \right)$
\STATE  $w_t^i = {{\mathop{\rm argmin}\nolimits} _w}\frac{1}{2}\left\| {p_t^i - w} \right\|_2^2 + {\lambda _t}{\left\| w \right\|_1}$ \\
\STATE  predict ${\widehat y}_t^i$
\STATE  receive $y_t^i$ and obtain $f_t^i(w_t^i ): = \ell (w_t^i,x_t^i,y_t^i)$
\STATE $\theta_{t + 1}^i = \sum\nolimits_j {{a_{ij}}\widetilde{\theta} {{_t^j} }}  - {\alpha _t}g_t^i$, where $g_t^i = \nabla f_t^i(w_t^i)$
\STATE  broadcast to neighbors: ${\widetilde{\theta} _{t + 1}^i} = \theta _{t + 1}^i + \delta _t^i$
\ENDFOR
\ENDFOR
\end{algorithmic}
\end{algorithm}

\section{Privacy Analysis}
As mentioned,  exploiting differential privacy (DL) protects the privacy while guarantees the usability of social data. In step 11 of Algorithm 1, $\theta $ is the parameter exchanged, to which we add a random noise. The added noise leads to the perturbation of $\theta $, so someone else cannot mine individual privacy according to an exact parameter. To recall, DL is defined mathematically in Definition 1, which aims at weakening the significantly difference between $\mathcal A\left( X \right)$ and $\mathcal A\left( {X'}\right)$.
Only satisfying the  inequality (4), can we ensure the privacy of social data in each  data center.

\subsection{Adding Noise}
Since  we add noise to mask the difference of two datasets differing at most in one point, the  sensitivity should be known.  Dwork \cite{dwork2006} proposed that the magnitude of the noise depends on the largest change that a single entry in data source could have on the output of Algorithm 1; this quantity is referred to as the \emph{sensitivity} of the algorithm.  The sensitivity of Algorithm 1 in defined.

\textbf{Definition 2 (Sensitivity).} Based on Definition 1, for any $\mathcal{X}$ and $\mathcal{{X'}}$, which differ in exactly one entry, we define the  sensitivity of Algorithm 1 at $t$-th round as
\begin{align}
{\rm{S}}({\rm{t}}){\rm{ = }}\mathop {\sup }\limits_{{\cal X},{\cal X}'} {\left\| {{\cal A}\left( {\cal X} \right){\rm{ - }}{\cal A}\left( {{\cal X}'} \right)} \right\|_{\rm{1}}}.
\end{align}

\textbf{Lemma 1.} Under Assumption 1, if the $L_1$-sensitivity of the parameter $\theta$  is computed as (5), we obtain
 \begin{equation}{\rm{S}}(t) \le 2{\alpha _t\sqrt n L},
 \end{equation}
where $n$ denotes the dimensionality of the vectors.
\begin{proof} 
See Algorithm 1,  $\theta$ is the exchanged parameter and added with the noise $ \delta$. According to Definition 1, we have \vspace{-.5em}\[ \left\| {\mathcal{A\left( X \right)}- \mathcal{A\left( {X'} \right)}} \right\|_1=\left\| {\theta_t^i -{\theta_t^i}'} \right\|_1.\]
Assuming that the only differenct social data comes at time $t$, we have
\vspace{-.5em}\[\theta _{t + 1}^i = \sum\nolimits_j {{a_{ij}}\widetilde {\theta _t^i}}  - {\alpha _t}g_t^i\ and \ \
{\theta _{t + 1}^i}' = \sum\nolimits_j {{a_{ij}}\widetilde {\theta _t^i}}  - {\alpha _t}{g_t^i}',\]
where $ (x_t^i,y_t^i)$ and $ ({x_t^i}',{y_t^i}')$ lead to $g_t^i$ and ${g_t^i}'$ due to Step 9 and 10 in Algorithm 1.

Then, we have
\begin{align}
\notag&\left\| {\theta_{t+1}^i -{\theta_{t+1}^i}'} \right\|_1\\
\notag& = \left\| { (\sum\nolimits_j {{a_{ij}}\widetilde {\theta _t^i}}- {\alpha _t}g_t^i)  -( \sum\nolimits_j {{a_{ij}}\widetilde {\theta _t^i}}  - {\alpha _t}{g_t^i}')}\right\|_1\\
\notag& \le {\alpha _t}\sqrt n \left( {{{\left\| {g_{t }^i} \right\|}_2} + {{\left\| {g_{t}^i}' \right\|}_2}} \right)\\
& \le 2{\alpha _{t}}\sqrt nL.
\end{align}

By Definition 2, we know  \vspace{-.5em}\begin{equation}\notag{\rm{S}}(t) \le \left\| {\theta_t^i -{\theta_t^i}'} \right\|_1.
 \end{equation}

Finally, combining (5) and (7), we obtain (6).
\end{proof}

 We determine the  magnitude of the noise as follows.
 $\sigma \in {\mathbb{R}^n}$ is a Laplace random noise vector drawn independently according to the density function:
\begin{equation}
Lap\left( {x|\mu } \right) = \frac{1}{{2\mu }}\exp \left( { - \frac{{\left| x \right|}}{\mu }} \right),
\end{equation}
where $\mu  = {{S\left( t \right)} \mathord{\left/ {\vphantom {{S\left( t \right)}\epsilon}} \right. \kern-\nulldelimiterspace} \epsilon}$. After this, we use $Lap\left( \mu  \right)$ to denote the Laplace distribution.

\subsection{Guaranteeing $\epsilon$-Differentially Private}
In our system model, as an  independent cloud node,  each data center should protect the privacy at every moment. If there is a data center invaded by a malicious user, this ``bad kid''  is able to get some information about other users' social data stored in  other data center across the network. Hence, every data transmitted should be processed by DL (i.e., satisfy (4)). Recalling from  Fig.1, we add random noise to every communication in the data center network.

 Having described the method and  magnitude of adding noise, we next prove how to guarantee $\epsilon$-differentially private for $\theta$. First, we demonstrate the privacy preserving at each time $t$.

\textbf{Lemma 2.} At the $t$-th round, the $i$-th cloud node's output of $\mathcal{A}$, $\widetilde \theta_t^i$, is $\epsilon$-differentially private.
\vspace{-.8em}
\begin{proof}
Let $\widetilde \theta_t^i = \theta_t^i + \sigma _t^i$ and $\widetilde {\theta}{_t^i}' = \theta{_t^i}' + \sigma _t^i$, then
by the definition of differential privacy (see Definition 1), $\widetilde \theta_t^i$ is $\epsilon$-differentially private if
\vspace{-.8em}\begin{equation}\Pr [\widetilde \theta_t^i ] \le {e^\epsilon}\Pr [\widetilde {\theta}{_t^i}' ].
\end{equation}
 We have
\begin{align}
\notag\frac{{\Pr \left( {\widetilde \theta_t^i} \right)}}{{\Pr \left( {\widetilde {\theta}{_t^i}'} \right)}}& = \prod\limits_{j = 1}^n {\left( {\frac{{\exp \left( { - \frac{{\epsilon\left| {\theta_t^i[j] - \theta[j]} \right|}}{{S\left( t \right)}}} \right)}}{{\exp \left( { - \frac{{\epsilon\left| {{\theta_t^i}'[j] -\theta[j]} \right|}}{{S\left( t \right)}}} \right)}}} \right)} \\
\notag& \le \prod\limits_{j = 1}^n {\exp \left( {\frac{{\epsilon\left| {{\theta_t^i}'[j] -\theta_t^i[j]} \right|}}{{S\left( t \right)}}} \right)} \\
\notag &= \exp \left( {\frac{{\epsilon{{\left\| {{\theta_t^i}' -\theta_t^i} \right\|}_1}}}{{S\left( t \right)}}} \right)\le \exp \left( \epsilon \right),
\end{align}
where the first inequality follows from the triangle inequality,  and the last  inequality follows from Lemma 1.
\end{proof}

McSherry \cite{mcsherry2009} has proposed that the privacy guarantee does not degrade across rounds as the samples used in the rounds are disjoint.  Obviously, our system model is an online processing website, where the social data is flowing. We dynamically serve the users with favorite recommendations due to users' recent social behavior. Hence,  during the $T$ rounds of our Algorithm 1, the  social data are disjoint. As  Algorithm 1 runs, the privacy guarantee will not degrade. Then we obtain the following theorem.

\textbf{Theorem 1 (Parallel Composition).} On the basis of Definition 1 and 3, under Assumption 1 and Lemma 2, our algorithm is $\epsilon$-differentially private.

For details of  proof of  Theorem 1, readers are advised to \cite{mcsherry2009}.

\section{utility analysis}
We have mentioned the notion regret, which is used to estimate the utility of online learning algorithms. The regret of  our online learning algorithm represents a  sum of  mistakes, which are made by all data centers during the learning and predicting process. When social websites conduct personalized recommendations (e.g., songs, videos and news etc.) for users, not all recommendations make sense for individuals. But we wish that with the system working and more social data being learnt, the predictions used for recommending become  more accurate. That means the regret should have  an upper bound.
Therefore,  lower regret bounds indicates  better and faster  distributed online learning algorithms. Firstly, we give the definition of  ``regret''.

\textbf{Definition 3.} We propose Algorithm 1 for social websites over data center networks. Then, we measure the regret of the algorithm as
\vspace{-.6em}\begin{eqnarray}
 {R} = \sum\limits_{t = 1}^T {\sum\limits_{i = 1}^m {f_t^i} } (w_t) - \mathop {\min }\limits_{w \in W} \sum\limits_{t = 1}^T {\sum\limits_{i = 1}^m {f_t^i} } (w),
\end{eqnarray}
where ${w_t} = \frac{1}{m}\sum\nolimits_i {w_t^i} $, denotes the average of $m$ parameters of all data centers at time $t$.
Hence, ${R}$ is computed with respect to an average of  $m$ parameters $w_t^j$, which approximately estimates the actual performance of the whole system.

For analyzing the regret $R$ of Algorithm 1, we firstly present a special lemma.

\textbf{Lemma 3.} Let $\varphi _t$ be  $\beta$-strongly convex functions, which have the norms ${\left\|  \cdot  \right\|_{{\varphi _t}}}$
 and dual norms $\left\|  \cdot  \right\|_{{\varphi _t}}^ * $.  When Algorithm 1 keeps running, we have  the following inequality\vspace{-.6em}
\begin{align}
\notag&\sum\limits_{t = 1}^T {\sum\limits_{i = 1}^m {{{\left( {{w_t} - w} \right)}^T}} } {g_t}\\
 \notag&\le {{m{\varphi _T}\left( w \right)} \mathord{\left/
 {\vphantom {{m{\varphi _T}\left( w \right)} {{\alpha _t}}}} \right.
 \kern-\nulldelimiterspace} {{\alpha _t}}} + \frac{1}{{{\alpha _t}}}\sum\limits_{t = 1}^T {\sum\limits_{i = 1}^m {\left[ {\varphi _t^ * \left( {{\theta _t}} \right) - \varphi _{t - 1}^ * \left( {{\theta _t}} \right)} \right.} } \\
 &\quad+ \left. {\frac{{{\alpha ^2}_t}}{{2\beta }}\left\| {{g_t}} \right\|_2^2 + {\alpha _t}{{\left\| {{\delta _t}} \right\|}_2} + {\lambda _t}{{\left\| {{g_t}} \right\|}_1}} \right].
\end{align}
 \begin{proof}
We define $\Phi _t^i = \varphi _t^ * \left( {{\theta _{t + 1}}} \right) - \varphi _{t-1}^ * \left( {{\theta _t}} \right)$, where ${\theta _t} = \frac{1}{m}\sum\nolimits_i {\theta _t^i} $.\vspace{-.6em}
\begin{align}
\notag{\theta _{t + 1}} &= \frac{1}{m}\sum\nolimits_i {\theta _{t + 1}^i}  = \frac{1}{m}\sum\nolimits_i {\left( {\sum\nolimits_j {{a_{ij}}\widetilde {\theta _t^j}}  - {\alpha _t}g_t^i} \right)} \\
 \notag&= \frac{1}{m}\sum\nolimits_j {\left( {\sum\nolimits_i {{a_{ij}}} } \right)\widetilde {\theta _t^j} - \frac{1}{m}\sum\nolimits_i {\alpha _tg_t^i} } \\
\notag& = \frac{1}{m}\sum\nolimits_j {\widetilde {\theta _t^j}}  - \frac{1}{m}\sum\nolimits_i {\alpha _tg_t^i} \\
\notag& = \frac{1}{m}\sum\nolimits_j {\left( {\theta _t^j + \delta _t^j} \right)}  - \frac{1}{m}\sum\nolimits_i {\alpha _tg_t^i} \\
 \notag&= {\theta _t}{\rm{ + }}\frac{1}{m}\sum\nolimits_j {\left( {\delta _t^j - \alpha _tg_t^j} \right)} \\
\, & = {\theta _t} + {\delta _t} - {\alpha _tg_t},
\end{align}
where intuitively ${\delta _t} = \frac{1}{m}\sum\nolimits_j {\delta _t^j} $ and ${g_t} = \frac{1}{m}\sum\nolimits_j {g_t^j} $.

First, according to Fenchel-Young inequality, we have
\begin{align}
\notag\sum\limits_{t = 1}^T {\Phi _t^i}  &= \varphi _T^ * \left( {{\theta _{T + 1}}} \right) - \varphi _0^ * \left( {{\theta _1}} \right) = \varphi _T^ * \left( {{\theta _{T + 1}}} \right)\\
 &\ge {w^T}{\theta _{T + 1}} - {\varphi _T}\left( w \right).
\end{align}
\vspace{-.6em}
Then, \begin{align}
\notag&\Phi _t^i = \varphi _t^ * \left( {{\theta _{t + 1}}} \right) - \varphi _t^ * \left( {{\theta _t}} \right) + \varphi _t^ * \left( {{\theta _t}} \right) - \varphi _{t - 1}^ * \left( {{\theta _t}} \right)\\
 &\le \varphi _t^ * \left( {{\theta _t}} \right) - \varphi _{t - 1}^ * \left( {{\theta _t}} \right) - {\alpha _t}{{\nabla \varphi _t^ * \left( {{\theta _t}} \right)}^T}{g_t} + \frac{{{\alpha _t^2}}}{{2\beta }}\left\| {{g_t}} \right\|_2^2 + {\alpha _t}{\left\| {{\delta _t}} \right\|_2}.
\end{align}

Combining (14) and (15), summing over $T=1,...T$  and $i=1,...,m$ , we get
\begin{small} \begin{align}
\notag&\sum\limits_{t = 1}^T {\sum\limits_{i = 1}^m {{\alpha _t}{{\left( {\nabla \varphi _t^ * \left( {{\theta _t}} \right) - w} \right)}^T}} } {g_t}\\
 &\le m{\varphi _T}\left( w \right) + \sum\limits_{t = 1}^T {\sum\limits_{i = 1}^m {\left[ {\varphi _t^ * \left( {{\theta _t}} \right) - \varphi _{t - 1}^ * \left( {{\theta _t}} \right) + \frac{{{\alpha ^2}_t}}{{2\beta }}\left\| {{g_t}} \right\|_2^2 + {\alpha _t}{{\left\| {{\delta _t}} \right\|}_2}} \right]} }.
\end{align}
\end{small} 
According to Lemma 1 of  Wang et al.\cite{wang2015}, we know
\begin{equation}{w_t}^T{g_t} \le \nabla \varphi _t^ * {\left( {{\theta _t}} \right)^T}{g_t} + {\lambda _t}{\left\| {{g_t}} \right\|_1} \end{equation}

Finally, using (16) and (17), we obtain (12).
\end{proof}

Based on Lemma 3, we easily have the regret bound of our system model.

\textbf{Theorem 2.} We propose Algorithm 1 for social big data computing over data center networks.  Under Assumption 1 and 2, we define regret function as (11). Set ${\varphi _t}\left( w \right) = \frac{1}{2}\left\| w \right\|_2^2$, which is $1$-strongly convex. Let ${\lambda _t} = {\alpha _t}\lambda $, then the regret bound is
\begin{equation}R \le R\sqrt {\left( {L + \lambda } \right)mTL}  + \frac{{2\sqrt 2 {m^2}nTL}}{\epsilon}\left( {\sqrt T  - \frac{1}{2}} \right)\end{equation}

\begin{proof}
For  convex functions, we know that
\[f_t^i\left( {{w_t}} \right) - f_t^i\left( w \right) \le {\left( {{w_t} - w} \right)^T}{g_t}.\]

Intuitively, due to (11) and (12), we obtain
\begin{align}
\notag R &\le {{m{\varphi _T}\left( w \right)} \mathord{\left/
 {\vphantom {{m{\varphi _T}\left( w \right)} {{\alpha _t}}}} \right.
 \kern-\nulldelimiterspace} {{\alpha _t}}} + \frac{1}{{{\alpha _t}}}\sum\limits_{t = 1}^T {\sum\limits_{i = 1}^m {\left[ {\varphi _t^ * \left( {{\theta _t}} \right) - \varphi _{t - 1}^ * \left( {{\theta _t}} \right)} \right.} } \\
 &\quad\,+ \left. {\frac{{{\alpha ^2}_t}}{{2\beta }}\left\| {{g_t}} \right\|_2^2 + {\alpha _t}{{\left\| {{\delta _t}} \right\|}_2} + {\lambda _t}{{\left\| {{g_t}} \right\|}_1}} \right]
\end{align}

Since ${\varphi _t}\left( w \right) = \frac{1}{2}\left\| w \right\|_2^2$,  we have $\varphi _t^ * \left( {{\theta _t}} \right) - \varphi _{t - 1}^ * \left( {{\theta _t}} \right) = 0$ and $\beta=1$.\vspace{-.6em}

\[R \le \underbrace {\frac{{\frac{1}{2}\left\| w \right\|_2^2}}{{{\alpha _t}}} + \frac{{mT{L^2}{\alpha _t}}}{2} + {\alpha _t}\lambda mTL}_{S1} + \underbrace {mT\sum\limits_{t = 1}^T {\sum\limits_{i = 1}^m {{{\left\| {{\delta _t}} \right\|}_2}} } }_{S2},\]
where $\left\| {{g_t}} \right\| \le L$ is defined previously.

We first compute $S1$:
setting ${\alpha _t} = \frac{{{{\left\| w \right\|}_2}}}{{2\sqrt {\left( {L + \lambda } \right)mTL} }}$, we have 
\begin{equation}S1 \le R\sqrt {\left( {L + \lambda } \right)mTL} \sim O\left( {\sqrt T } \right), \end{equation} where $R$ is defined in Assumption 2.

Then, for $S2$, we have
 \begin{align}
\notag &E\left[ {\sum\limits_{t = 1}^T {\sum\limits_{i = 1}^m {\left\| {\sigma _t^i} \right\|} } } \right] \le \frac{{2\sqrt 2mnL }}{\epsilon}\left( {\sqrt T  - \frac{1}{2}} \right),\\
& S2 \le \frac{{2\sqrt 2 {m^2}nTL}}{\epsilon}\left( {\sqrt T  - \frac{1}{2}} \right)\sim O\left( {\sqrt T } \right).
\end{align}

Combining (20) and (21), we obtain (18).
\end{proof}


\begin{figure}
\begin{minipage}[t]{0.5\linewidth}
\centering
\includegraphics[width=1.9in]{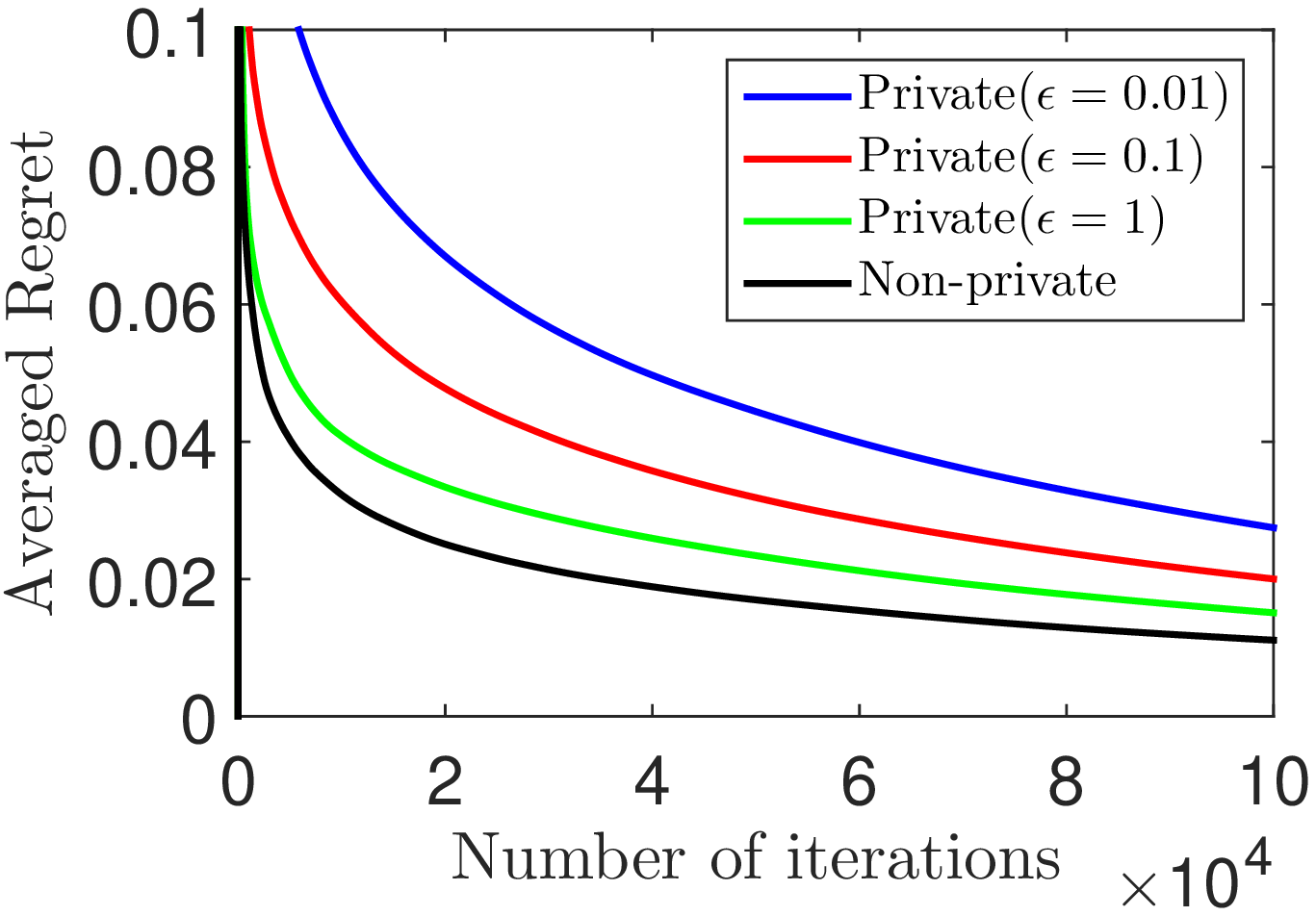}\vspace{-.9em}
\caption{ Nodes=64 and Sparsity=52.3{\%}}
\label{fig:side:a}
\end{minipage}%
\begin{minipage}[t]{0.6\linewidth}
\centering
\includegraphics[width=1.9in]{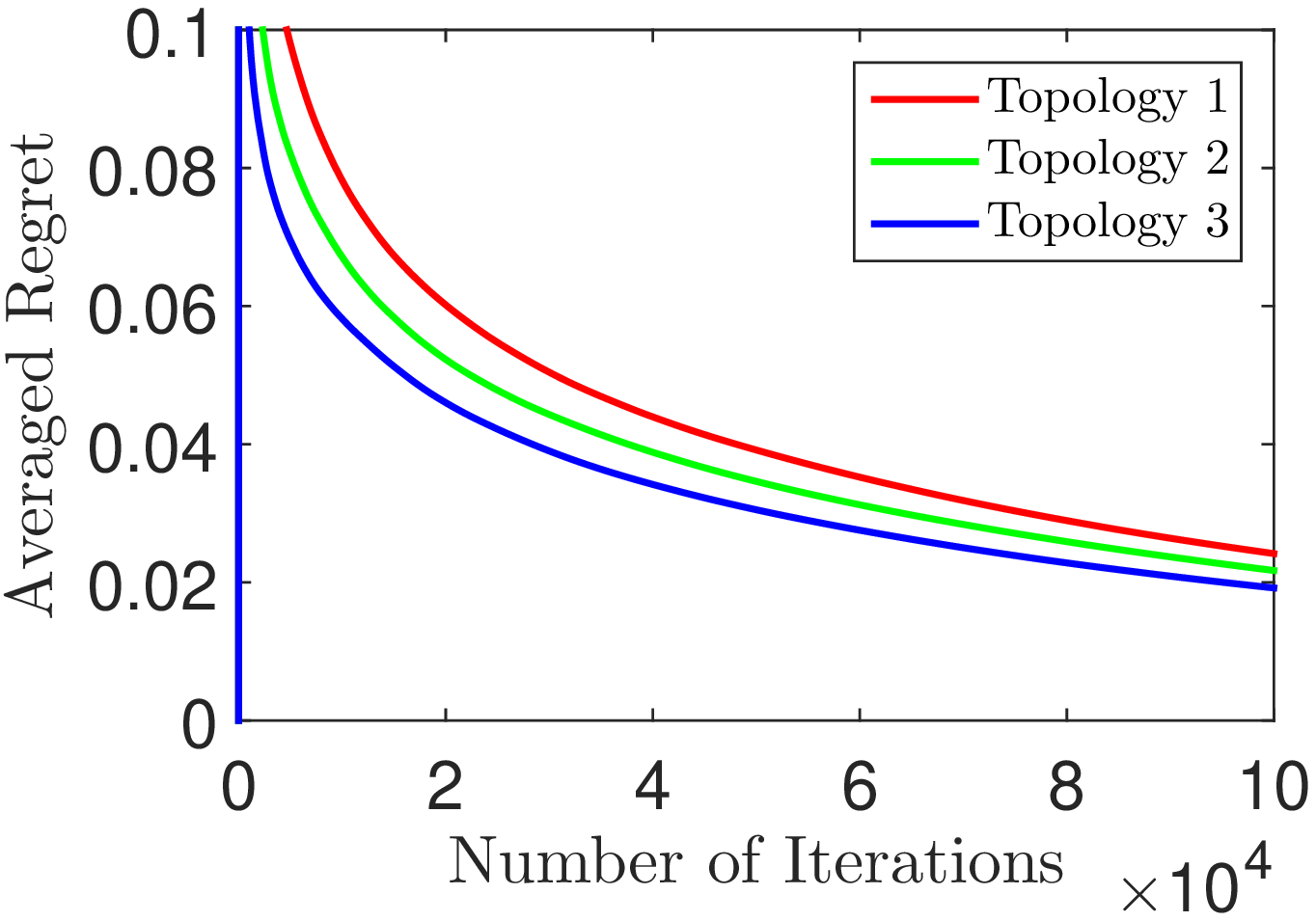}\vspace{-.9em}
\caption{Nodes=64 and $\epsilon=0.1$}
\label{fig:side:b}
\end{minipage}
\begin{minipage}[t]{0.5\linewidth}
\centering
\includegraphics[width=1.78in]{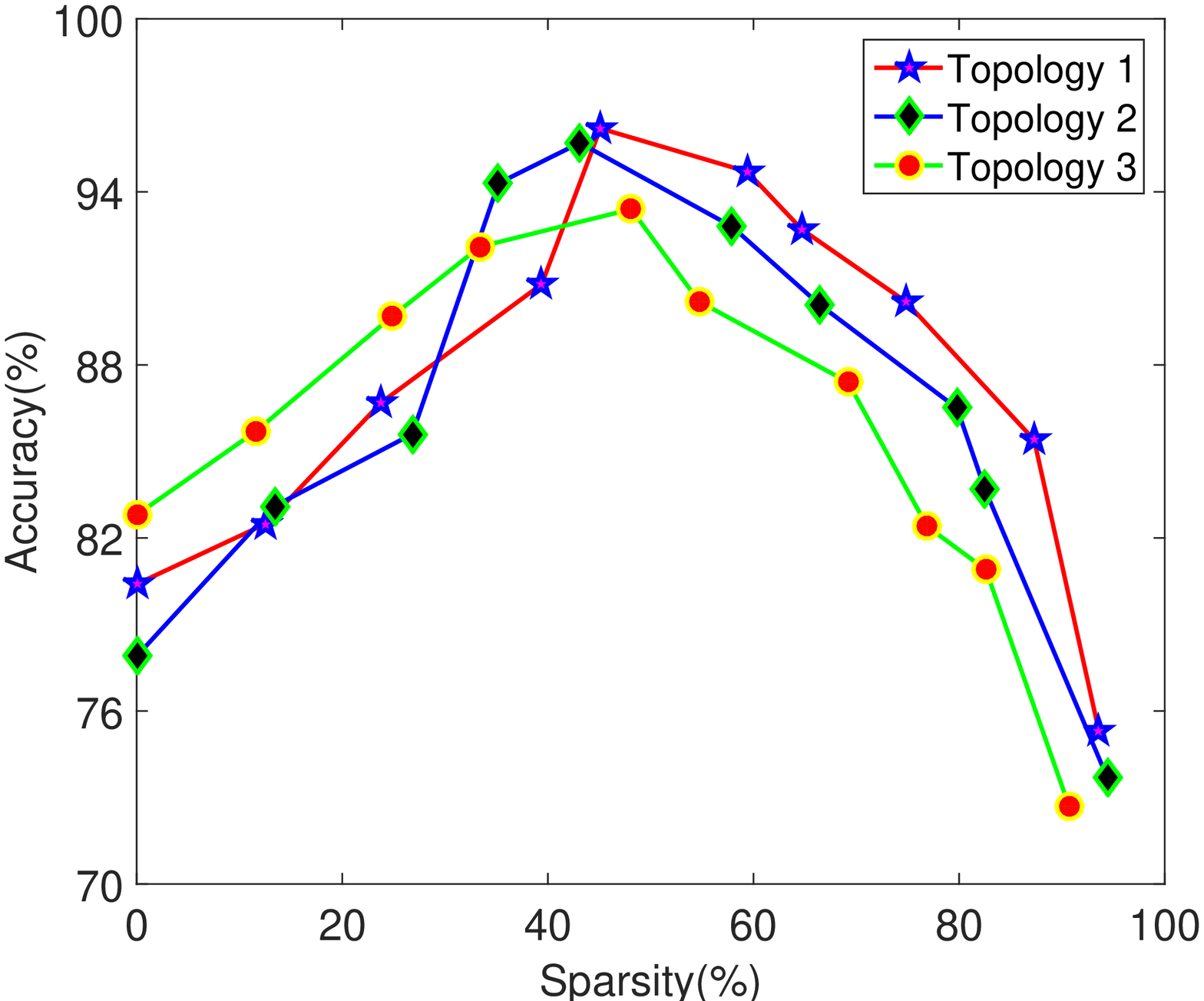}
\caption{Nodes=64 and $\epsilon=0.1$}
\label{fig:side:c}
\end{minipage}%
\begin{minipage}[t]{0.6\linewidth}
\centering
\includegraphics[width=1.8in]{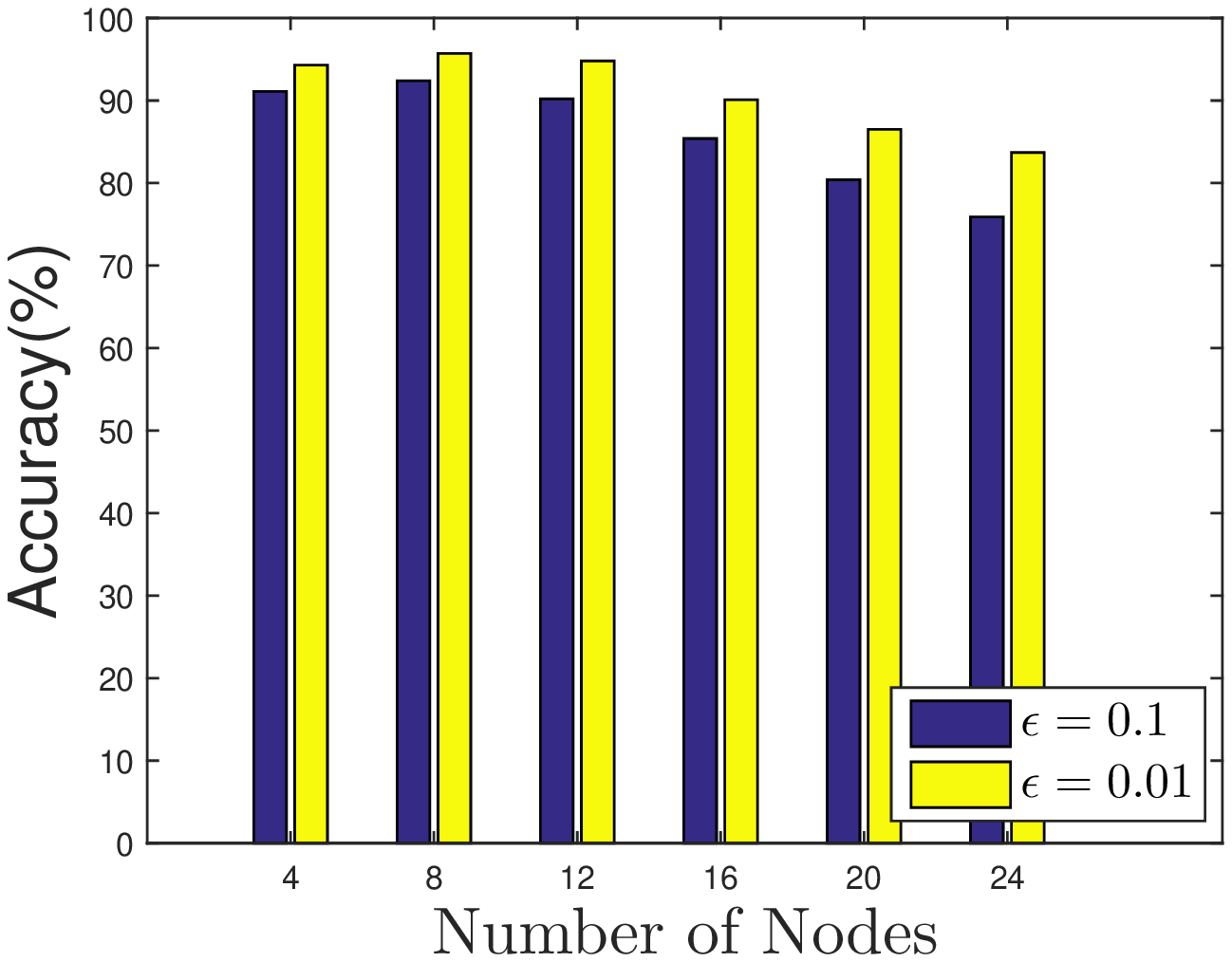}
\caption{ Sparsity=64.5{\%}}
\label{fig:side:d}
\end{minipage}
\end{figure}

According to Theorem 2, the regret bound becomes the classical square root regret  $O(\sqrt T )$ \cite{zinkevich2003}, which means less mistakes are made in social recommendations  as the algorithm runs. This result demonstrates that our private online learning algorithm for the social system makes sense. Further, due to (18), we find: 1) a higher privacy level $\epsilon$ can enhance the regret bound; 2) the number of data centers gets more, the regret bound become higher; 3) the communication matrix $a_{ij}$ seems not to affect the bound, but we think it may affect the convergence. All the observations will be simulated in the following numerical experiments.

\section{simulations}
In this section, we conduct four simulations. The first one is to study the privacy and predictive performance trade-offs. The second one is to find whether the topology of social networks has a big influence on the performance. The third one is to  study the sparsity and performance trade-offs. The final one is to analyze the performance trade-offs between the number of data centers and accuracy.
All the simulations are operated on real  large-scale and high-dimension  social data. 

 For our implementations, we have  the hinge loss $f_t^i\left( w \right) = \max \left( {1 - y_t^i\left\langle {w,x_t^i} \right\rangle } \right)$, where $\left\{ {\left( {x_t^i,y_t^i} \right) \in {\mathbb{R}^n} \times \left\{ { \pm 1} \right\}} \right\}$, are the data available only to the $i$-th data center. For powers of persuasion, we use $100,000$ social data to experiment and the dimensionality of data is $10,000$. Since the tested data are  real social data, we should pretreat the data. Each dimension in vectors is normalized into a certain numerical interval. Each data point is  labeled with a value into $\left[ {0,1} \right]$ according to its classification attribute. For the simulated model, we design it as Fig.1. A few computing nodes are distributed randomly. Each node is only connected with its adjacent nodes. Everytime information exchanging is perturbed with Laplace noise. 
All the experiments were conducted on a distributed model designed by Hadoop under Linux (with 8 CPU cores, 64GB memory).
 
In Fig.2, the regret bound of the non-private algorithm has the lowest regret as expected and it  shows that  the  regret gets closer to the non-private regret as its privacy preservation is weaker. The higher privacy level of $\epsilon$ leads to the more regret.  Fig.3 demonstrates that different topologies make no significant difference on the utility of the algorithm.
 Fig.4 indicates that an appropriate sparsity can have the best performance and  other lower or higher sparsity would  lead to a worse performance. Specifically, inducing sparsity can enhance the accuracy, obtaining nearly $18{\%}$ more than the non-sparse computing does. Fig.5 studies the performance with respect to the number of data center nodes. More centers can have a little  decline (as much as $4\%$ per 4 nodes) in the accuracy.

\section{conclusion}
Internet has come into Big Data era. Social networks are faced with massive data to handle. Faced with these challenges, we proposed a private distributed online learning algorithm for social big data over data center networks. We demonstrated that higher privacy level leads to weaker utility of the system and the appropriate sparsity enhances the performance of online learning for high-dimension data.
Furthermore, there must exist delay  in social networks, which we did not consider. Hence, we hope that online learning with delay can be presented in the future work.

\section*{Acknowledgment}
This research is supported by National Science Foundation of China with Grant 61401169.

\end{document}